\documentclass[3p,preprint,12pt]{elsarticle}
\usepackage{amsmath,amsfonts,amssymb}

\usepackage{graphicx}
\usepackage{color}

\def\sech{{\rm sech}}
\def\arctanh{{\rm tanh^{-1}}}

\journal{Physics Letters A}

\begin{document}
\begin{frontmatter}

\title{Particle and wave dynamics of nonlocal solitons in external potentials}

\author[uoa]{G. N. Koutsokostas}
\address[uoa]{Department of Physics, National and Kapodistrian University of Athens, Panepistimiopolis, Zografos, Athens 15784, Greece}

\author[uoa]{I. Moseley}

\author[uoi]{T. P. Horikis}
\address[uoi]{Department of Mathematics, University of Ioannina, Ioannina 45110, Greece}

\author[uoa]{D. J. Frantzeskakis}

\begin{abstract}

We study nonlocal bright solitons subject to external 
spatially nonuniform potentials. If the potential is slowly varying on 
the soliton scale, we derive analytical soliton solutions behaving 
like Newtonian particles. 
If the potential has the form of an attractive delta-like point defect, 
we identify different dynamical regimes, defined by the relative strength of the 
nonlocality and the point defect. In these regimes, the soliton can be 
trapped at the defect's location, via a nonlinear resonance with a defect mode --
which is found analytically-- reflected by or transmitted through the defect, 
featuring a wave behavior. Our analytical predictions are 
corroborated by results of direct numerical simulations. 

\end{abstract}

\begin{highlights}

\item This work explores the dynamics of nonlocal bright solitons in slowly varying and point defect external potentials. Analytical and numerical methods confirm soliton behavior similar to Newtonian motion, including parabolic trajectories and harmonic oscillations in linear and parabolic potentials, respectively.

\item The asymptotic regimes based on defect strength relative to nonlocal nonlinearity are identified. Solitons exhibit wave dynamics (reflection/transmission) when defects dominate, form defect modes when defect and nonlinearity are balanced, and exhibit behavior similar to the local NLS equation when nonlinearity dominates.

\item Our methodology offers a rather universal perspective on the particle-wave duality of 
nonlocal solitons and is able to capture the analytical soliton solutions of the relative problem.

\end{highlights}


\end{frontmatter}

\section{Introduction}

We consider the following one-dimensional (1D) normalized nonlocal  
nonlinear Schr{\"o}dinger equation for the complex field $u(x,t)$:
\begin{eqnarray}
i u_t + \frac{1}{2}u_{xx} + nu = V(x)u,
\label{nlnls} 
\end{eqnarray}
where subscripts denote partial derivatives, $V(x)$ is a real external potential, 
and the function $n(x,t)$ is given by:
\begin{eqnarray}
n(x,t)=\int_{-\infty}^{+\infty} R(x-x')|u(x',t)|^2 dx'. 
\label{nlnls2}
\end{eqnarray}
Here, the kernel $R(x)$ is a real function, which describes the nonlocal response 
of the medium; notice that if the response function is singular, i.e., 
$R(x) = \delta(x)$ (with $\delta(x)$ being the Dirac $\delta$ function) then 
Eq.~(\ref{nlnls})reduces to the local nonlinear Schr{\"o}dinger (NLS) equations, or 
Gross-Pitaevskii (in the context of Bose-Einstein condensates (BECs) \cite{bec}) 
equation. We assume that Eqs.~(\ref{nlnls})-(\ref{nlnls2}) are supplemented with 
homogeneous boundary conditions at infinity, and consider localized in space 
solutions, i.e., solitons.

The above model may find applications in the dynamics of optical beams 
in nonlocal nonlinear media. In such a case, $u(x,t)$ represents the complex 
electric field envelope, $t$ denotes the propagation distance, 
while the function $n(x,t)$ may play the role of: 
(a) the nonlinear change of the refractive index (that depends on the light 
intensity $I=|u|^2$) in thermal media \cite{th1,th2}, (b) the relative electron 
temperature perturbation in plasmas \cite{pl1,pl2}, and (c) the perturbation 
of the optical director angle from its static value, due to the presence of the 
light field, in nematic liquid crystals \cite{lq1,lq2}. The considered nonlocal 
NLS model is also relevant to the physics of quasi-1D dipolar BECs \cite{dbec}, 
with $u(x,t)$ representing the macroscopic wavefunction, and $V(x)$ being the 
external trapping potential. The latter, may also be relevant to optics, with 
$V(x)$ typically accounting for the spatial modulation of the linear part of 
the refractive index \cite{kivsharagr}. 

There exist various studies on the dynamics of solitons in external 
potentials. A central result in this context is that, for potentials slowly-varying 
on the soliton scale, the soliton center $x_0$ features a {\it particle-like} nature, 
obeying the following Newtonian equation of motion:
\begin{equation}
\frac{d^2 x_0}{dt^2}=-\frac{dV(x_0)}{dx_0},
\label{nem}
\end{equation}   
which is consistent with the Ehrenfest theorem of quantum mechanics \cite{landau}. 
In the case of a parabolic potential, $V(x) = (1/2)\Omega^2 x^2$, Eq.~(\ref{nem}) 
implies that the soliton performs a harmonic oscillation of frequency $\Omega$. 
This result was first found for solitons of the local NLS~\cite{kos,moura} 
(see also Refs.~\cite{hulet,hern} and the review \cite{Nonlin} for BECs), and later 
was also shown in the case of nonlocal solitons for the case of thermal 
media~\cite{segchr}, for liquid crystals~\cite{lq1,lq2}, and generalized to 
the case of topological defects~\cite{natphot}. Notice that one of the first 
explored type of spatial solitons in liquid crystals was actually propagating 
in an external potential written by the applied voltage \cite{assoe}. 
Newtonian equations of motion, similar to~(\ref{nem}), were also derived for solitons 
in trapped dipolar BECs by means of a variational approximation \cite{abdul}. 
Furthermore, more recently, the use of external 
potentials in nonlocal NLS systems have been proposed as a means to control and 
manipulate nonlinear waves \cite{c1,tph,c2}. 

On the other hand, soliton dynamics have also been studied in the case where 
the soliton's spatial width is much smaller than the characteristic spatial 
scale of the external potential \cite{bishop}. The most prominent and relevant 
example in such settings corresponds to the case where the potential takes the 
form of a point impurity (or defect) of strength $\epsilon$, namely 
$V(x) = -\epsilon \delta(x)$. Pertinent soliton-impurity collisions where first 
studied in the context of the local NLS equation in various works ---see, e.g.,  
Refs.~\cite{kos,cao,roy,jeremy}--- and later for the nonlocal setting of dipolar BECs 
\cite{abdul}. It was shown that, if the point impurity is 
attractive ($\epsilon>0$), and if the soliton velocity is small while $\epsilon$ is 
sufficiently large, then solitons can be trapped at the impurity location; i.e., 
a nonlinear resonance of the soliton with a {\it defect mode} (the nonlinear analogue 
of the bound state occurring in the linear Schr{\"o}dinger equation \cite{landau}) 
takes place. This trapping, can also be accompanied by the splitting of the 
incoming soliton into a reflected and a transmitted part. Hence, in this  
case, where the potential has the form of a point impurity, the soliton follows 
a {\it wave} behavior upon its interaction with the impurity.  
Generally, the soliton may behave either as particles or as waves 
in external potentials, thus featuring a {\it particle-wave duality} 
(see discussion in Refs.~\cite{kos,assdual}), depending on the relative widths 
of the soliton and the external potential. This behavior was observed also 
experimentally in nonlocal media, namely in lead glasses featuring a 
thermal nonlinearity \cite{segchr} and in nematic liquid crystals 
across boundaries \cite{natphys}.

Here, in the framework of Eq.~(\ref{nlnls}), we consider bright sech$^2$-shaped 
nonlocal solitons (see Ref.~\cite{krolexsol} for this type of solitons), and 
study their particle and wave dynamics. We introduce an 
analytical approach, combined with asymptotic and numerical techniques,    
which complements previous work in the contexts of both local and nonlocal 
NLS models; for the latter, different settings and methods have been considered~\cite{lq1,lq2,segchr,natphot,abdul,natphys}. Exploiting 
the existence of exact analytical 
soliton solutions of the nonlocal NLS~(\ref{nlnls}) for a specific form 
of $R(x)$, we present results 
which, although qualitatively similar to existing ones, refer to a wide class of 
nonlocal systems, including thermal media, plasmas and nematic liquid crystals,   
as mentioned above. Thus, our work offers a novel, quite general methodology 
and provides a rather universal picture 
for the nonlocal solitons' particle-wave duality. A brief description of our 
findings and the organization of our presentation is as follows.   
 
First, in Sec.~II we revisit the derivation of the nonlocal soliton solutions 
of Eq.~(\ref{nlnls}) for $V(x)=0$. Then, we study soliton dynamics in the presence  
of the potential $V(x)$ and investigate, in particular, the following cases: 
\begin{itemize}
\item[(i)] $V(x)$ is varying slowly on the soliton scale (Sec.~III),  
\item[(ii)] the spatial scale of $V(x)$ is much smaller than that of the 
soliton; for the latter case, we specifically assume that $V(x)=-\epsilon \delta(x)$ 
(Sec.~IV).
\end{itemize}
To study case (i), we devise an analytical methodology, relying on the use of the 
hydrodynamic form of the model and an approach resembling the adiabatic 
perturbation theory for solitons \cite{kar2,kivmal}, to derive approximate 
soliton solutions in the presence of $V(x)$. We show that these solitons 
behave as particles, following  Newtonian dynamics, with the evolution of their 
center obeying Eq.~(\ref{nem}). 
For case (ii), we identify three different asymptotic regimes, pertinent to 
the nonlocality and the external potential scale competition: in particular, these 
regimes are defined by the relative strength of the defect and the nonlocality parameter $d$. Then, for each of them, we  
investigate soliton-defect interactions, and find the following: 
\begin{itemize}
\item[(a)] If the defect dominates nonlinearity, the soliton features wave 
dynamics, being reflected by or transmitted through the defect, closely following 
dynamics governed by the linear Schr{\"o}dinger equation (see Refs.~\cite{cao,abdul} 
for the local and nonlocal NLS setting, respectively).

\item[(b)] If the roles of the defect and the nonlinearity are of equal importance, 
we analytically determine a defect mode solution, and using numerical simulations,  
we find that the soliton-defect collision may lead to either complete trapping of the 
soliton by the defect or partial reflection and transmission, accompanied by 
partial trapping of the soliton. 

\item[(c)] If the nonlinearity dominates the effect of the impurity, we 
find that, similarly to the case of the local NLS, fast solitons are transmitted 
through the defect, while slow solitons are captured in the vicinity of the defect, 
and perform long-lived oscillations. 
\end{itemize}
Finally, in Sec.V we summarize our findings and discuss directions for future work.

\section{Nonlocal bright solitons}

We now focus on a specific form of the nonlocality, such that the kernel 
$R(x)$ is a real, positive definite, localized and symmetric function, obeying 
the normalization condition $\int_{-\infty}^{+\infty}R(x)dx =1$. 
We consider in particular the following form of the kernel, which is 
relevant to the evolution of an electromagnetic beam in a nonlocal nonlinear medium 
\cite{th1,th2,pl1,pl2,lq1,lq2} (see also \cite{job,dum,bam1} and 
references therein):
\begin{equation}
R(x)=\frac{1}{2d}\exp\left(-\frac{|x|}{d} \right).
\label{Rofx}
\end{equation}
Here, $d > 0$ is a spatial scale accounting for the degree of nonlocality 
(for $d \rightarrow 0$ the function $R(x)$ becomes singular, and Eq.~(\ref{nlnls})  
reduces to the NLS with a local nonlinearity). In this case, 
using Fourier transforms,  
it can be shown 
(see, e.g., Ref.~\cite{g1jpa}) that Eq.~(\ref{nlnls}) can be expressed as 
the following system of partial differential equations (PDEs): 
\begin{eqnarray}
&&i u_t + \frac{1}{2} u_{xx} + nu = V(x)u,
\label{1} \\
&&d^2 n_{xx} - n + |u|^2=0.
\label{2}
\end{eqnarray}

We consider at first the case where the potential is absent, i.e., $V(x)=0$, with our 
aim being in discussing the possible form of solitons solutions that can be 
supported by Eqs.~(\ref{1})-(\ref{2}). Such nonlocal solitons can typically be 
found in approximate form (see, e.g., Refs.~\cite{appr1,appr2}), featuring a shape 
that depends on the degree of nonlocality, namely ranging from a sech-profile 
(in the local case, with $d\rightarrow 0$) to a Gaussian profile (in the highly 
nonlocal case, with $d\rightarrow \infty$). Nevertheless, for a fixed and finite 
value of the nonlocality parameter $d$ (which, as will be seen, controls the 
soliton amplitude ---and power), exact analytical sech$^2$-shaped soliton 
solutions of Eqs.~(\ref{1})-(\ref{2}) are possible ---see Ref.~\cite{krolexsol} and 
references therein. 

Let us now revisit the derivation of the aforementioned exact soliton solutions  
which, for simplicity, will first be derived in a stationary form. We introduce 
the ansatz:
\begin{equation}
u=u_0(x)\exp[i\omega_0(t+\sigma_0)], \quad n = n_0(x),
\label{ans}
\end{equation}
where $u_0(x)$ is an unknown real function, $\omega_0$ is the unknown frequency 
of the solution, while $\sigma_0$ is an arbitrary real parameter representing the
initial phase of the soliton. Substituting Eqs.~(\ref{ans}) into
Eqs.~(\ref{1})-(\ref{2}) (for $V(x)=0$), we arrive at the following system 
of ordinary differential equations (ODEs):
\begin{eqnarray}
&&u_0'' -2\omega_0 u_0 + 2 u_0 n_0=0,
\label{ode1}\\
&&d^2 n_0'' - n_{0}+ u_0^2=0,
\label{ode2}
\end{eqnarray}
with primes denoting differentiation with respect to $x$. Then, observing that if: 
\begin{equation}
n_0=\frac{1}{\sqrt{2}d}u_0, \quad {\rm and} \quad \omega_0 =\frac{1}{2d^2},
\label{thq}
\end{equation}
then the system~(\ref{ode1})-(\ref{ode2}) reduces to a single ODE:
\begin{equation}
u_0'' - 2\omega_0 u_0 + \sqrt{\frac{2}{d}} u_0^2 =0.
\end{equation}
Then, the exact solution  
$u_0(x)= \frac{3}{2\sqrt{2}d}\sech^2\left(\frac{x-x_0}{2d}\right)$ of the latter 
(with $x_0$ being the initial location, or the center of the pulse),   
gives rise to the following nonlocal bright soliton solutions of 
Eqs.~(\ref{1})-(\ref{2}):
\begin{eqnarray}
u(x,t)&=&\frac{3}{2\sqrt{2}d}
\sech^{2}\left[\frac{1}{2d}(x-x_0)\right]e^{i\omega_0 (t + \sigma_0)},
%\exp[i\omega_0 (t + \sigma_0)],
\label{4}\\
n(x,t)&=&\frac{3}{4d^2} \sech^{2}\left[\frac{1}{2d}(x-x_0)\right].
\label{5}
\end{eqnarray}

Apart from stationary solutions, traveling soliton solutions exist as well, 
and can be constructed by means of a Galilean boost. Indeed, it is straightforward 
to see that if $u_s(x,t)$, $n_s(x,t)$ is a stationary solution of 
Eqs.~(\ref{1})-(\ref{2}) for $V(x)=0$, then a traveling wave solution of these 
equations is of the form:
\begin{eqnarray}
u(x,t)&=&u_s(x-kt,t)\exp\left[i\left(kx-\frac{1}{2}k^2 t\right)\right], 
\label{G1}\\
n(x,t)&=&n_s(x-kt,t),
\label{G2}
\end{eqnarray}
where $k$ is a free parameter characterizing the velocity, wavenumber and frequency 
of the traveling solution. Hence, using Eqs.~(\ref{G1})-(\ref{G2}) and 
Eqs.~(\ref{4})-(\ref{5}), we find that the traveling soliton solutions of 
Eqs.~(\ref{1})-(\ref{2}), for $V(x)=0$, are:
\begin{eqnarray}
\!\!\!\!\!\!\!
u(x,t)&=&\frac{3}{2\sqrt{2}d}
\sech^{2}\left[\frac{1}{2d}(x-kt-x_0)\right]e^{i\phi(x,t))},
\label{4b} \\
\!\!\!\!\!\!\!
\phi(x,t)&=&kx-\frac{1}{2}\left(k^2-\frac{1}{d^2}\right)t+\sigma_0,
\label{5b} \\
\!\!\!\!\!\!\!
n(x,t)&=&\frac{3}{4d^2} \sech^{2}\left[\frac{1}{2d}(x-kt-x_0)\right].
\label{6b}
\end{eqnarray}

\section{Solitons in slowly-varying potentials}

We now consider the full problem where the potential is present, and consider 
at first the case of slowly-varying potentials. In particular, we will study  
the case where the soliton width, characterized by the nonlocality parameter $d$ [see 
Eqs.~(\ref{4b})-(\ref{6b})] is such that: 
\begin{equation}
d \ll W, 
\label{dW}
\end{equation}
where $W$ is the characteristic scale of the inhomogeneity of $V(x)$.   
We wish to show that, in this case, the 
problem admits a bright soliton solution, of a form similar to that given in 
Eqs.~(\ref{4b})-(\ref{6b}), but with the soliton's center, wavenumber and phase 
being functions of time; the explicit form of these functions, together with 
the soliton shape, will be determined below in a self-consistent manner. 
Notice that our approach is reminiscent to the adiabatic approximation in the
perturbation theory for solitons \cite{kar2,kivmal}, which is commonly used in 
problems involving perturbed NLS models.

To proceed, we separate the real and imaginary parts in Eq.~(\ref{1}) using the 
ansatz $u(x,t)=f(x,t)\exp[i\theta(x,t)]$, where $f(x,t)$ 
and $\theta(x,t)$ are real functions. This way, we obtain from 
Eqs.~(\ref{1})-(\ref{2}) the following system of PDEs:
\begin{eqnarray}
&&\left(f^2\right)_t+\left(f^2\theta_x\right)_x =0,
\label{h1} \\
&&f_{xx}-\left[2\theta_t +\theta_x^2+2V(x) \right]f +2nf=0,
\label{h2} \\
&&d^2n_{xx}-n+f^2=0.
\label{h3}
\end{eqnarray}
Notice that, in fact, Eqs.~(\ref{h1}) and (\ref{h2}) constitute the hydrodynamic form 
of Eq.~(\ref{1}), with (\ref{h1}) being the continuity equation (conservation of mass) 
and (\ref{h2}) having the form of an Euler equation (conservation of momentum). 
Next, to further simplify the above system, we assume that the functions $f$ and 
$\theta$ are of the form:
\begin{eqnarray}
f=f(\xi), \quad \xi\equiv x-X_0(t), \quad \theta=K(t)x-\varphi_0(t),
\label{solans}
\end{eqnarray}
where $X_0(t)$ represents the location of the maximum of the function $f$, while 
$K(t)$ and $\varphi_0(t)$ denote, respectively, the wavenumber and time-dependent 
phase of the solution; all the above are unknown functions, to be determined. Then, 
substituting Eq.~(\ref{solans}) into Eqs.~(\ref{h1})-(\ref{h3}), we find the 
following. First, Eq.~(\ref{h1}) leads to the result:
\begin{equation}
K(t)=\dot{X}_0,
\label{Koft}
\end{equation}
with overdot denoting differentiation with respect to $t$. Second, Eq.~(\ref{h2})
becomes:
\begin{equation}
f_{\xi \xi}-2 \Omega_0 f + 2nf=0,
\label{h2b}
\end{equation}
where the function $\Omega_0(x,t)$ is given by:
\begin{equation}
\Omega_0(x,t)= -\dot{\varphi}_0 +\ddot{X_0}x +\frac{1}{2}\dot{X}_0^2 + V(x).
\label{Omofx}
\end{equation}
We now observe that the system of Eqs.~(\ref{h2b}) and (\ref{h3}) has a form similar 
to that of Eqs.~(\ref{ode1})-(\ref{ode2}), but with the important difference that 
Eq.~(\ref{h2b}) features the coefficient $\Omega_0$ which depends both on space $x$ 
and time $t$. This fact indicates that, under certain conditions, soliton solutions 
similar to those given in Eqs.~(\ref{4b})-(\ref{6b}) may also be supported by  
Eqs.~(\ref{h2b}) and (\ref{h3}), i.e., even in the presence of the external potential. 

To identify such conditions, first we notice that the spatial dependence of this 
coefficient may be suppressed upon assuming that the 
potential $V(x)$ varies slowly over the width of the density of the solution. 
To see this, we Taylor expand the potential around $X_0$, namely: 
\begin{equation}
V(x)=V(X_0) + \frac{dV}{dX_0}(x-X_0) 
+ \frac{1}{2}\frac{d^2V}{dX_0^2}(x-X_0)^2 +\cdots,
\nonumber
\end{equation}
where $dV/dX_0=dV/dx\big|_{x=X_0}$ and so on. Then, for a slowly varying 
potential [see Eq.~(\ref{dW})], we may keep in the Taylor 
expansion solely the linear terms in $(x-X_0)$. Substitution of these terms into 
Eq.~(\ref{Omofx}) leads to the following expression of $\Omega_0$:
\begin{eqnarray}
\Omega_0(x,t)
=-\dot{\varphi}_0 +\left[\ddot{X_0}+ \frac{dV}{dX_0}\right]x 
+\left[\frac{1}{2}\dot{X}_0^2 + V(X_0)\right]-X_0\frac{dV}{dX_0}.
\label{Omofx2}
\end{eqnarray}
It is now straightforward to see that as long as $\Omega_0$ depends only on time,  
the soliton center evolves according to the Newtonian equation of motion:
\begin{equation}
\ddot{X_0}=-\frac{dV}{dX_0},
\label{Neqmot}
\end{equation}
a result which is identical to the one corresponding to the local NLS [see 
Eq.~(\ref{nem})]. Thus, in the nonlocal problem under consideration, we again 
obtain a result consistent with the Ehrenfest theorem. 
Notice that the quantity in the second square bracket in the right-hand side 
of Eq.~(\ref{Omofx2}) represents the total energy $E_0$ of the pertinent dynamical 
system, namely:
\begin{equation}
E_0=\frac{1}{2}\dot{X}_0^2 + V(X_0) 
= \frac{1}{2}\dot{X}_0^2(0) + V(X_0(0)), %={\rm const.},
\label{nrg}
\end{equation}   
which remains constant for all $t$, taking a value determined by the initial velocity 
$\dot{X}_0(0)$ and initial location $X_0(0)$ of the solution. 

According to the above, we arrive at the following expression for %the function 
$\Omega_0=\Omega_0(t)$ in Eq.~(\ref{h2b}): 
\begin{equation}
\Omega_0=-\dot{\varphi}_0 + E_0 -X_0\frac{dV}{dX_0},
\label{Omofx3}
\end{equation} 
and impose the requirement $\Omega_0(t)={\rm const.}$   
This is in accordance to the line of the analysis of the previous Section 
[for Eqs.~(\ref{ode1}) and (\ref{ode2})], 
and leads to conditions rendering 
Eqs.~(\ref{h2b}) and (\ref{h3}) identical. Obviously, these conditions are similar 
to the ones in Eq.~(\ref{thq}), namely:
\begin{equation}
n=\frac{1}{\sqrt{2}d}f, \quad {\rm and} \quad \Omega_0 =\frac{1}{2d^2},
\label{thq2}
\end{equation}
where the second of the above equations can be used to determine the unknown 
function $\varphi_0(t)$. The latter, is given by:
\begin{equation}
\varphi_0(t)=\left(E_0-\frac{1}{2d^2}\right)t +\int_0^t X_0(s)\frac{dV}{dX_0}ds.
\label{varph}
\end{equation} 
To this end, summarizing the above results, we can arrive at the following 
expressions for the nonlocal bright solitons in external potentials:
\begin{eqnarray}
\!\!\!\!\!\!\!
u(x,t)&=&\frac{3}{2\sqrt{2}d}
\sech^{2}\left[\frac{1}{2d}\left(x-X_0(t)\right)\right]e^{i\varphi(x,t))},
\label{t1} \\
\!\!\!\!\!\!\!
\varphi(x,t)&=&\dot{X}_0(t)x-\varphi_0(t),
\label{t2} \\
\!\!\!\!\!\!\!
n(x,t)&=&\frac{3}{4d^2} \sech^{2}\left[\frac{1}{2d}\left(x-X_0(t)\right)\right],
\label{t3}
\end{eqnarray}
where the soliton center $X_0(t)$ obeys the Newtonian equation of 
motion~(\ref{Neqmot}), and $\varphi_0(t)$ is given by Eq.~(\ref{varph}). 

We proceed by presenting results of direct numerical simulations. 
In particular, we evolve 
Eqs.~(\ref{1})-(\ref{2}) using Eqs.~(\ref{t1})-(\ref{t3}) for $t=0$ as initial 
conditions, considering  
two different types of the external potential: a linear and a parabolic 
one, and compare the numerical results with our analytical predictions. 

\begin{figure}[tbp]
\centering
\includegraphics[height=5cm]{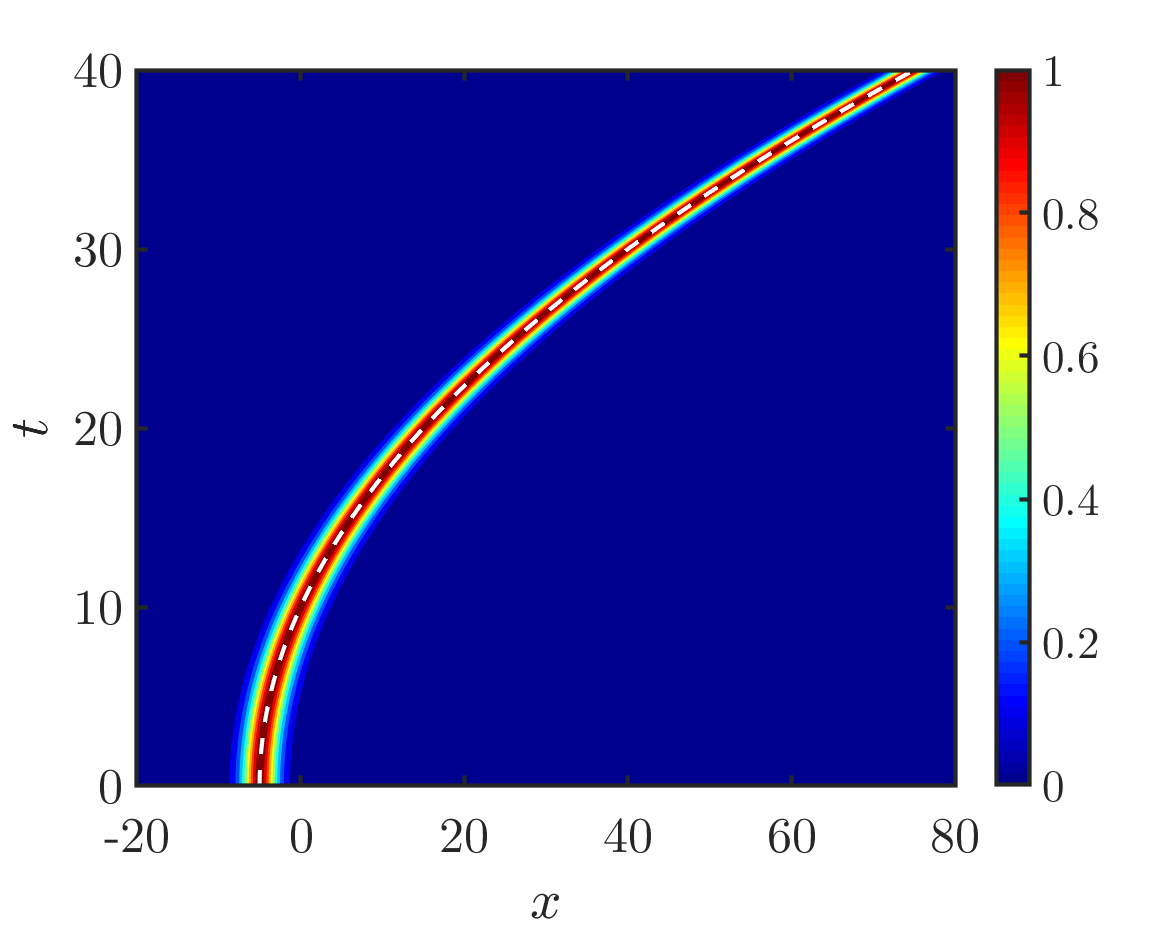}
\caption{Contour plot showing the evolution of the modulus $|u(x,t)|$ of 
a nonlocal bright soliton in the presence of the linear (gravitational-type) 
potential [Eq.~(\ref{linV})]. The dashed line depicts the soliton trajectory 
[Eq.~(\ref{X0linV})]. Parameter values are $a=-1$, $b=0$, $X_0(0)=-5$, 
and $d=1$.}
\label{fig1linV}
\end{figure}  

We start by noting the following. While %, generally, 
the expressions~(\ref{t1})-(\ref{t3}) for the soliton profiles are  
approximate, and are valid for arbitrary potentials varying slowly over the 
soliton scale, they become exact in the case of linear (gravitational-type) 
potentials, of the form: 
\begin{equation}
V(x)=ax+b, 
\label{linV}
\end{equation}
where $a, b \in \mathbb{R}$. In this case, a soliton subject to such a potential  features an accelerating motion: indeed, the soliton center is given by:
\begin{equation} 
X_0(t)=-\frac{1}{2}at^2 + X_0(0),
\label{X0linV}
\end{equation}
and the evolution of the soliton, as obtained from the numerical integration of 
Eqs.~(\ref{1})-(\ref{2}), is depicted in Fig.~\ref{fig1linV} for parameter 
values $a=-1$, $b=0$, $d=1$ and $X_0(0)=-5$. In this figure, shown is the 
contour plot of the soliton modulus, $|u(x,t)|$ in the $xt$-plane, with the 
dashed line depicting the analytical prediction of Eq.~(\ref{X0linV}). Observe 
that the soliton follows the analytically predicted dynamics, 
in excellent agreement with the numerical finding.

\begin{figure}[tbp]
\centering
\includegraphics[height=5cm]{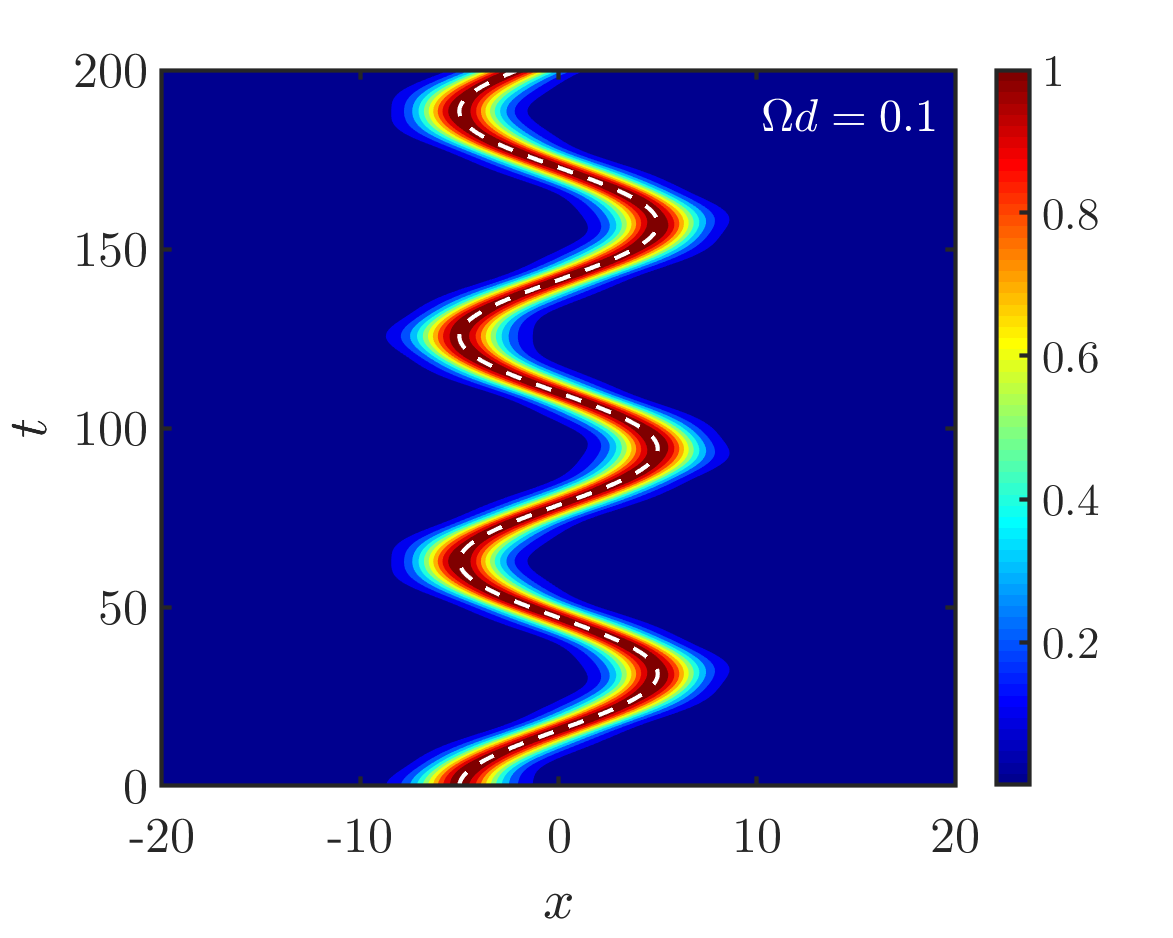}
\includegraphics[height=5cm]{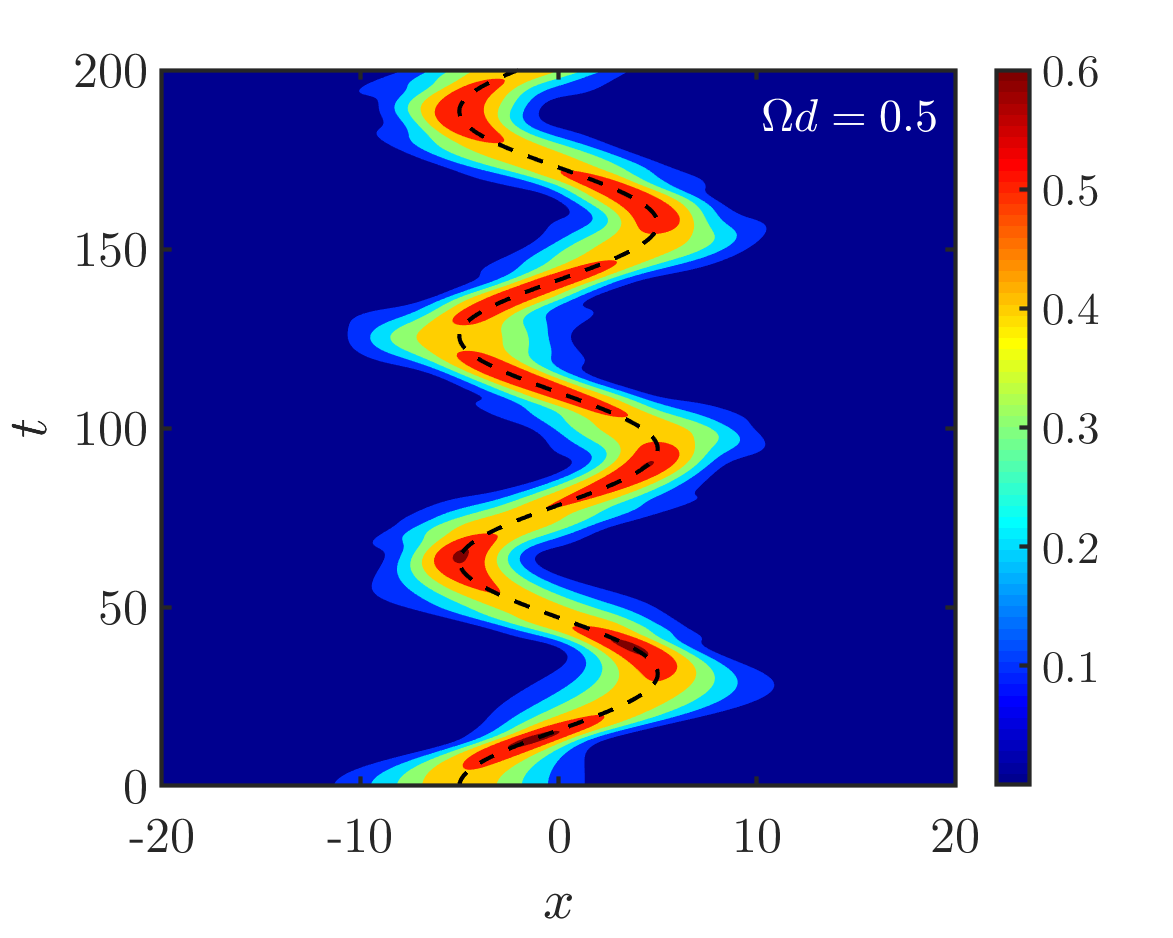}
\caption{Contour plots showing the evolution of the modulus $|u(x,t)|$ of 
a nonlocal bright soliton in the presence of the parabolic potential of 
Eq.~(\ref{parV}). The dashed lines depict the soliton trajectory 
given in Eq.~(\ref{X0parV}). Parameter values are $\Omega=0.1$, $X_0(0)=-5$, 
$\psi_0=0$, while $d=1$ or $d=5$ for the left and right panels, respectively.}
\label{fig2pV}
\end{figure} 

We now consider the case where the potential has the form of a parabolic trap, 
with a trap strength $\Omega$, namely: 
\begin{equation}
V(x)=\frac{1}{2}\Omega^2 x^2.
\label{parV}
\end{equation}
In this case, since the characteristic spatial scale of the potential is  
$W\sim \Omega^{-1}$, the requirement of Eq.~(\ref{dW}) is fulfilled for 
$\Omega d\ll 1$. Hence, as predicted, in this case the solitons  
perform harmonic oscillations of frequency $\Omega$, with the soliton 
center being given by:
\begin{equation}
X_0(t)= A_0 \cos(\Omega t+ \psi_0),  
\label{X0parV}
\end{equation}
where the amplitude of the oscillation $A_0$ and the initial phase $\psi_0$ are 
determined by the initial conditions. This scenario is confirmed by our numerical 
result shown in the left panel of Fig.~\ref{fig2pV}, where we have used 
$\Omega=0.1$ and $d=1$ and, hence, $\Omega d=0.1$ (other parameter values are  
$X_0(0)=-5$ and $\psi_0=0$). As seen in this figure, not only the trajectory 
of the soliton center follows the form of Eq.~(\ref{X0parV}) --see dashed line-- 
but also the functional form of the entire soliton structure evolves adiabatically. 

On the other hand, if the requirement of Eq.~(\ref{dW}) 
is violated, as, e.g., in the case example with $\Omega=0.1$ and $d=5$ 
(and thus $\Omega d=0.5$), then our approach ceases to be accurate. In particular, 
as shown in the right panel of Fig.~\ref{fig2pV}, the soliton is deformed while 
evolving in the trap: it features a periodic increase (decrease) in its width 
(amplitude), accompanied by emission of weak radiation, whenever its motion changes 
direction, and reorganizes its shape in the vicinity of the trap center. 
Nevertheless, the soliton still performs an oscillatory motion, with the 
frequency of the oscillation being equal to $\Omega$ (see dashed line in the figure). 
Notice that other cases with $\Omega d \sim 1$ were studied too (results not shown 
here); in such cases, a qualitatively similar behavior was found, but with the 
soliton featuring a stronger deformation, and emitting stronger radiation.

\section{Solitons and point defects}

In the previous Section, we analyzed the case where the soliton width $d$ 
is much smaller than the characteristic scale $W$ of the potential
[Eq.~(\ref{dW})]. Here, we will consider the case:
\begin{equation}
d \gg W, 
\label{dW2}
\end{equation}
with the most prominent example being the scenario of a point-like impurity, 
or defect, where the external potential has the form of a delta function: 
\begin{equation}
V(x)=-\epsilon\delta(x). 
\label{deltadef}
\end{equation}
Here, $\epsilon>0$ is a free parameter representing the strength of the defect. Notice 
that the minus sign corresponds to the case of an attractive impurity which, in the 
context of quantum mechanics, is known to give rise to a bound state 
\cite{landau}. On the other hand, in the nonlinear case, i.e., in the context 
of the NLS equation, the analogue of this bound state is the so-called 
``defect mode''. The latter, is a stationary nonlinear state located at the defect, 
and can be found in an exact analytical form \cite{roy}. 

We identify three different regimes, distinguished by the relative strength 
of the soliton amplitude ---which is determined by the nonlocality parameter, being 
$\propto 1/d$--- and the strength $\epsilon$ of the localized impurity. In particular, 
we will study the cases corresponding to: 
\begin{itemize}
\item[(A)] $\epsilon d \gg 1$ where the 
effect of impurity dominates nonlinearity (linear Schr\"{o}dinger regime); 

\item[(B)] $\epsilon d \sim 1$ (crossover regime), where the 
nonlinearity and impurity are of similar strengths; 

\item[(C)] $\epsilon d \ll 1$ (nonlinear regime), where the nonlinearity 
dominates the effect of the impurity.

\end{itemize}

\subsection{Linear Schr\"{o}dinger regime, $\epsilon d \gg 1$} 

We start with the regime $\epsilon d \gg 1$, which corresponds to the linear 
Schr\"{o}dinger limit: indeed, in this case, the nonlinearity term can be neglected 
(because $\lim_{d\rightarrow \infty} n(x,t) =0$) and the problem of 
Eqs.~(\ref{1})-(\ref{2}) can be approximated by the linear Schr\"{o}dinger equation 
with an attractive delta potential, namely:
\begin{eqnarray}
i u_t + \frac{1}{2} u_{xx}= -\epsilon \delta(x) u.
\label{ls}
\end{eqnarray}
As is well known \cite{landau}, the above equation features a normalized bound state, of energy $E_b=-\epsilon ^2/2$, which is of the form:
\begin{eqnarray}
u(x,t)=\sqrt{\epsilon} \exp\left({-\epsilon |x| -i\frac{\epsilon ^2}{2}t}\right),
\label{bns}
\end{eqnarray}
as well as a continuous spectrum consisting of scattering states, 
$\propto \exp[{i(kx-\omega t})]$, of wavenumber $k$ and frequency $\omega$. 
For the scattering states, the 
reflection and transmission coefficients, $b(k)$ and $a(k)=1+b(k)$ respectively, 
read (see, e.g., Ref.~\cite{cao}): $b(k)=-\epsilon/(\epsilon +ik)$, and $a(k)=ik/(\epsilon +ik)$; hence, 
\begin{equation}
|b(k)|^2=\frac{\epsilon^2}{\epsilon^2+k^2}, \quad 
|a(k)|^2=\frac{k^2}{\epsilon^2+k^2}, 
\label{TR}
\end{equation}
and it is noted that $|a(k)|^2+|b(k)|^2=1$.

The above simple analysis suggests the following. Consider a nonlocal soliton, 
of the form of Eq.~(\ref{4b}), initially located sufficiently far from the defect;  
this soliton, obviously, is an exact solution of the problem. Now assume that this 
structure moves towards the defect at $x=0$. In the considered regime of 
$\epsilon d \gg1$, since we may neglect the nonlinearity, the soliton of 
Eq.~(\ref{4b}) (at $t=0$) can be interpreted by its Fourier transform, meaning 
that each Fourier component of wavenumber $k$ will independently interact with 
the defect. In other words, the wavenumber $k$ in Eqs.~(\ref{TR}) can be 
treated as the free parameter of the moving nonlocal soliton and, thus, 
$|a(k)|^2$ and $|b(k)|^2$ for the soliton may be approximated by Eq.~(\ref{TR}). 
Furthermore, we may predict that total reflection (transmission) 
occurs in the limit $\epsilon^2 \gg k^2$ ($\epsilon^2 \ll k^2$). Outcomes 
involving both reflection and transmission are obviously possible too.   
 
\begin{figure}[tbp]
\centering
\includegraphics[height=4.3cm]{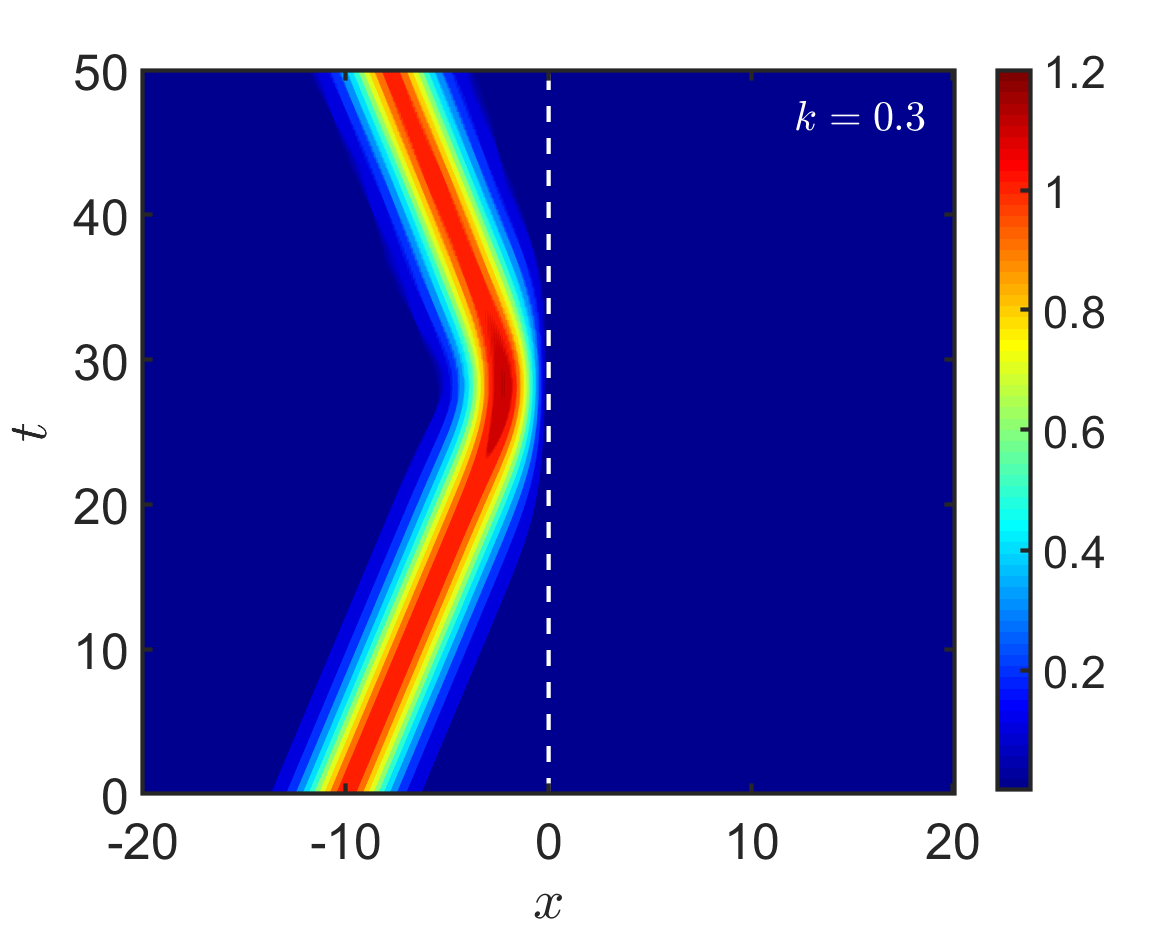}
\includegraphics[height=4.3cm]{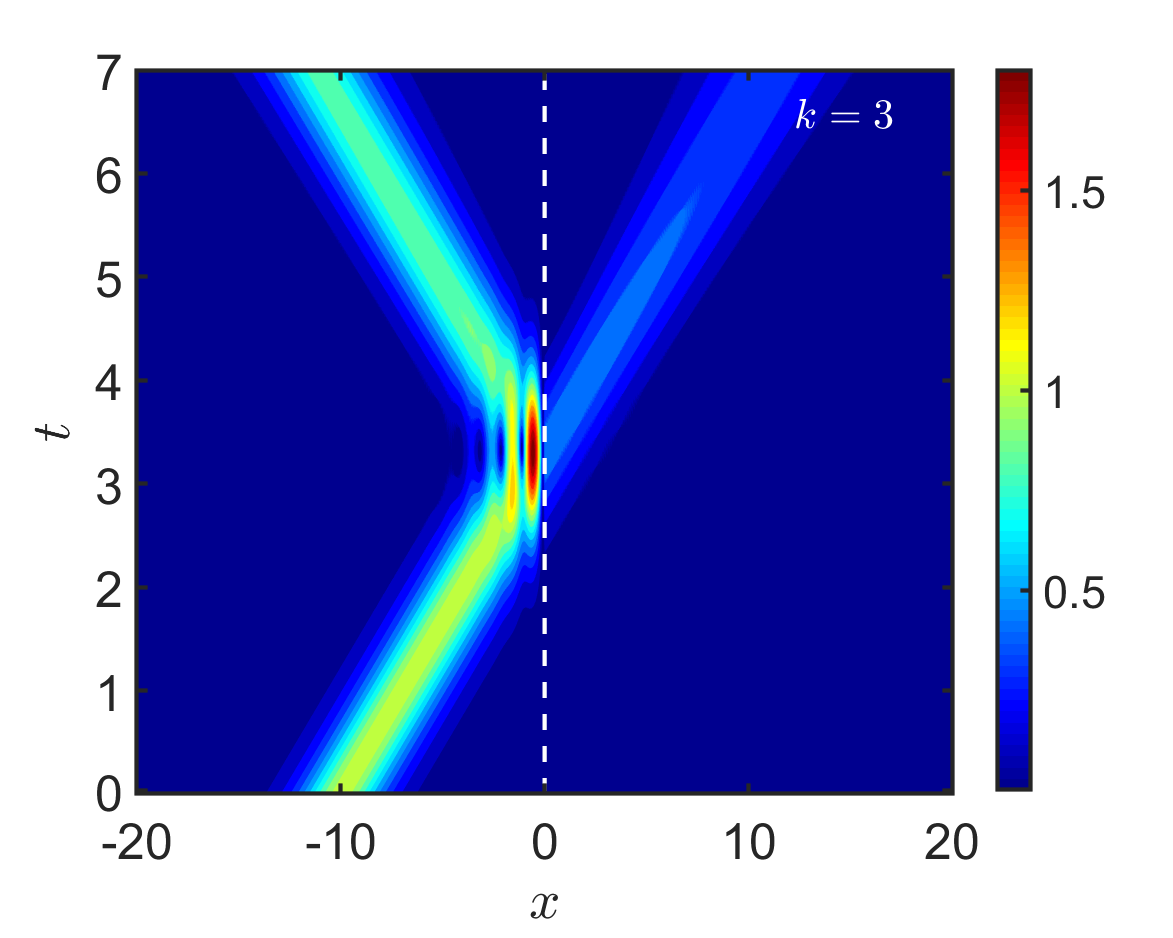}
\includegraphics[height=4.3cm]{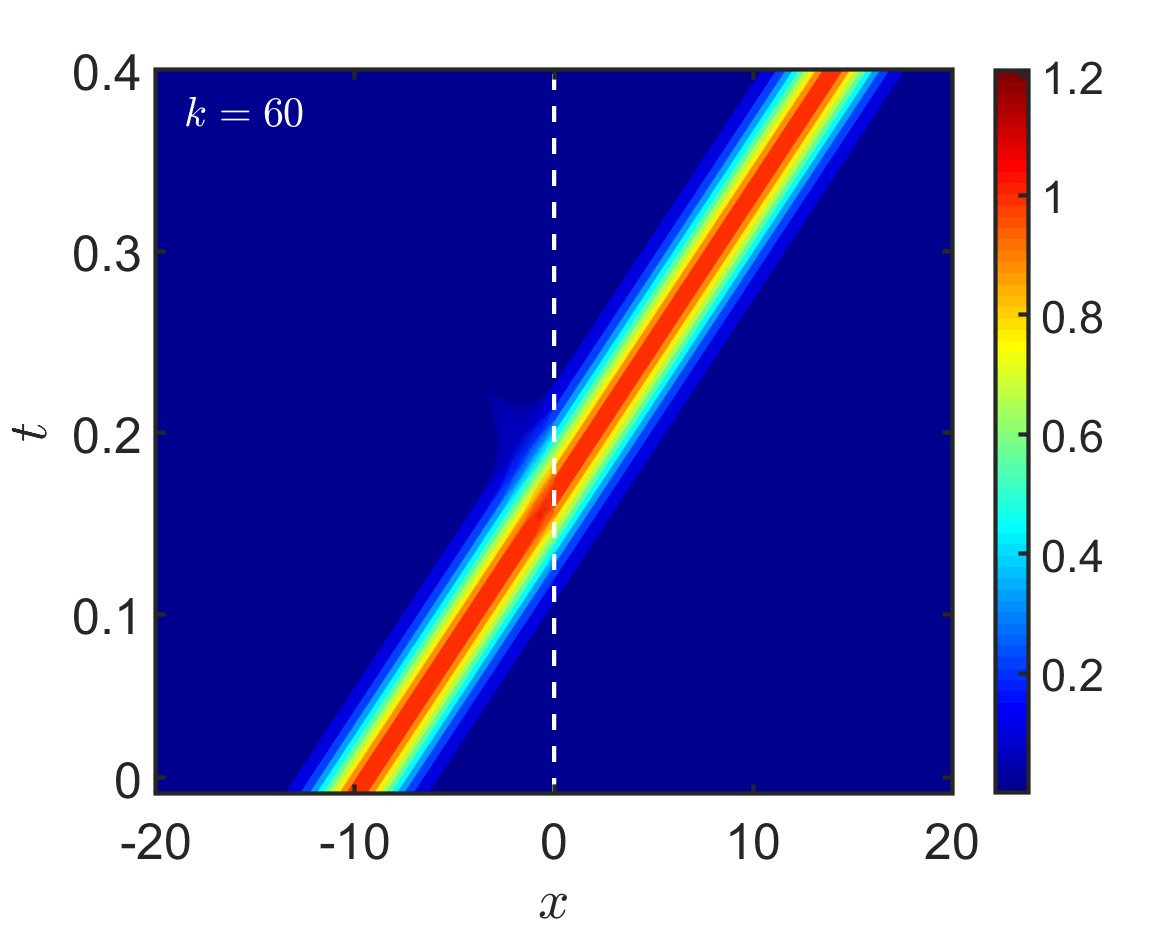}
\caption{Contour plots showing the evolution of the modulus $|u(x,t)|$ of 
a nonlocal bright soliton, initially placed at $x_0=-10$, for $k=0.3$ (left   
panel), $k=3$ (middle panel) and $k=60$ (right panel); other parameter values 
are $\epsilon = 5$ and $d=1$, corresponding to the linear Schr{\"o}dinger regime, 
$\epsilon d \gg 1$. 
Dashed (white) line depicts the location of the point defect. 
From top to bottom, observed are the scenarios of total reflection, partial 
reflection and partial transmission, and total 
transmission, as predicted in Eq.~(\ref{TR}). }
\label{figls1}
\end{figure}

Thus, in this regime, the soliton features a {\it wave} behavior 
(see discussion, e.g., in Refs.~\cite{kos,assdual} for the local and nonlocal NLS 
respectively), which is highlighted in Fig.~\ref{figls1}. In this figure, shown are 
contour plots of the field $|u(x,t)|$ in the $xt$-plane, as obtained by means 
of direct simulations; the dashed (white) lines depict the location of the defect. 
The initial condition in all cases was the soliton of 
Eq.~(\ref{4b}), initially located at $x_0=-10$. The strength of the defect and 
the nonlocality parameter were fixed to the values are $\epsilon = 5$ and $d=1$, 
while the soliton wavenumber (initial momentum) assumed the values $k=0.3$ (left  
panel), $k=3$ (middle panel) and $k=60$ (right panel). As predicted in our 
analysis above, the first (third) case corresponds to total reflection 
(transmission), while the second one leads to both reflection and transmission. 
Observe that, in this latter case, only a small part of the soliton is
captured by the defect, while large parts are reflected and transmitted, with 
the relevant amounts being predicted fairly well by Eqs.~(\ref{TR}); notice that a 
similar behavior was also found in Ref.~\cite{cao}, but for the local NLS model. 

\subsection{Crossover regime, $\epsilon d \sim 1$} 

\begin{figure}[!htb]
\centering
\includegraphics[height=6cm]{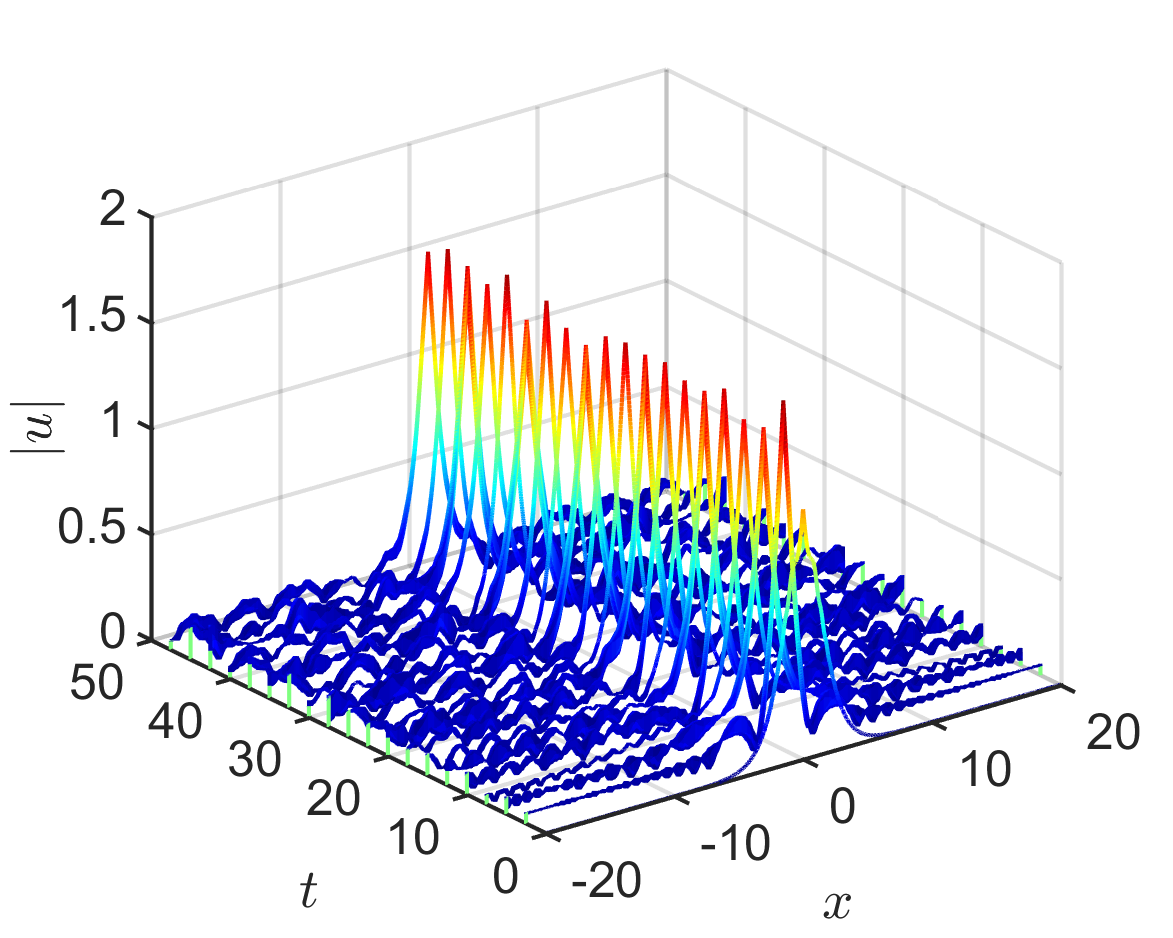}
\caption{A 3D plot showing the evolution of the modulus $|u(x,t)|$ of 
a stationary soliton, given by 
Eq.~(\ref{4}) with $d=1$, when initialized at the location of a point defect 
of amplitude $\epsilon=0.5$. Observe that the initial soliton emits radiation and eventually reorganizes itself into a defect mode.}
\label{figdm}
\end{figure} 

Next, we consider the crossover regime, with $\epsilon d \sim 1$, 
where the effects of the nonlinearity and the impurity play equally important 
roles in the dynamics. 
In this case, we may employ our analytical approach of the previous 
Section to determine, at first, the defect mode for the 
nonlocal problem under consideration. Our starting point is  
the system 
of Eqs.~(\ref{h1})-(\ref{h3}). As we are seeking for a stationary nonlinear 
state located at $x=0$, we assume that 
%it is natural to consider the case where 
$$(f^2)_t=0, \quad \theta_x=0, \quad \theta_t=\omega_0=\frac{1}{2d^2},$$ 
as suggested by the exact analytical form of the stationary 
nonlocal soliton [Eqs.~(\ref{4})-(\ref{5})]. Then, the continuity 
equation~(\ref{h1}) is automatically satisfied, while Eq.~(\ref{h2}) takes the form:
\begin{eqnarray}
f_{xx}-2\omega_0 f +2nf = 2\epsilon \delta(x)f.
\label{h2del}
\end{eqnarray}
Requiring continuity of the functions $f$ and $n$ at $x=0$, i.e., 
$f(0^{-})=f(0^{+})$ and $n(0^{-})=n(0^{+})$, we integrate the 
above equation from $x=0^-$ to $x=0^+$, and obtain the jump condition for the first 
spatial derivative of $f$, namely:
$\partial_{x}f(0^{+})-\partial_{x}f(0^{-})=2\epsilon f(0).$  
Then, integrating Eqs.~(\ref{h2del}) and (\ref{h3}) for $x<0$ and $x>0$, we arrive at 
the following solution, satisfying the above mentioned boundary conditions: 
\begin{eqnarray*}
f(x)=\frac{3}{2\sqrt{2}d}
\sech^{2}\left[\frac{1}{2d}\left(|x|-\chi_0 \right)\right], 
\quad n(x)=\frac{1}{\sqrt{2}d}f(x), 
\end{eqnarray*}
where $\chi_0=2d \arctanh(\epsilon d)$, with $0<\epsilon d<1$. 
To this end, the defect mode of the nonlocal problem 
---which is the nonlinear analogue of the bound state of Eq.~(\ref{bns}) occurring in 
the linear Schr{\"o}dinger limit--- is given by:
\begin{eqnarray}
\!\!\!\!\!\!\!\!\!
u(x,t)&=&\frac{3}{2\sqrt{2}d}
\sech^{2}\left[\frac{1}{2d}\left(|x|-\chi_0 \right)\right] 
\exp\left(i\frac{1}{2d^2}t \right),
\label{h2d} \\
\!\!\!\!\!\!\!\!\!
n(x,t)&=&\frac{4}{3d^2}
\sech^{2}\left[\frac{1}{2d}\left(|x|-\chi_0 \right)\right].
\label{h3d}
\end{eqnarray}
 
A simple way to observe the defect mode in the numerical simulations, is 
to initialize the stationary soliton~(\ref{4}) at $x=0$, i.e., at the 
location of the point defect. Indeed, in Fig.~\ref{figdm}, we observe that 
as the initial soliton evolves, it emits radiation and is eventually 
reorganized into the defect mode, which is characterized by a cusp occurring 
at $x=0$. Here, we have used the parameter 
values $\epsilon=0.5$ and $d=1$.     

\begin{figure}[tbp]
\centering
\includegraphics[height=5cm]{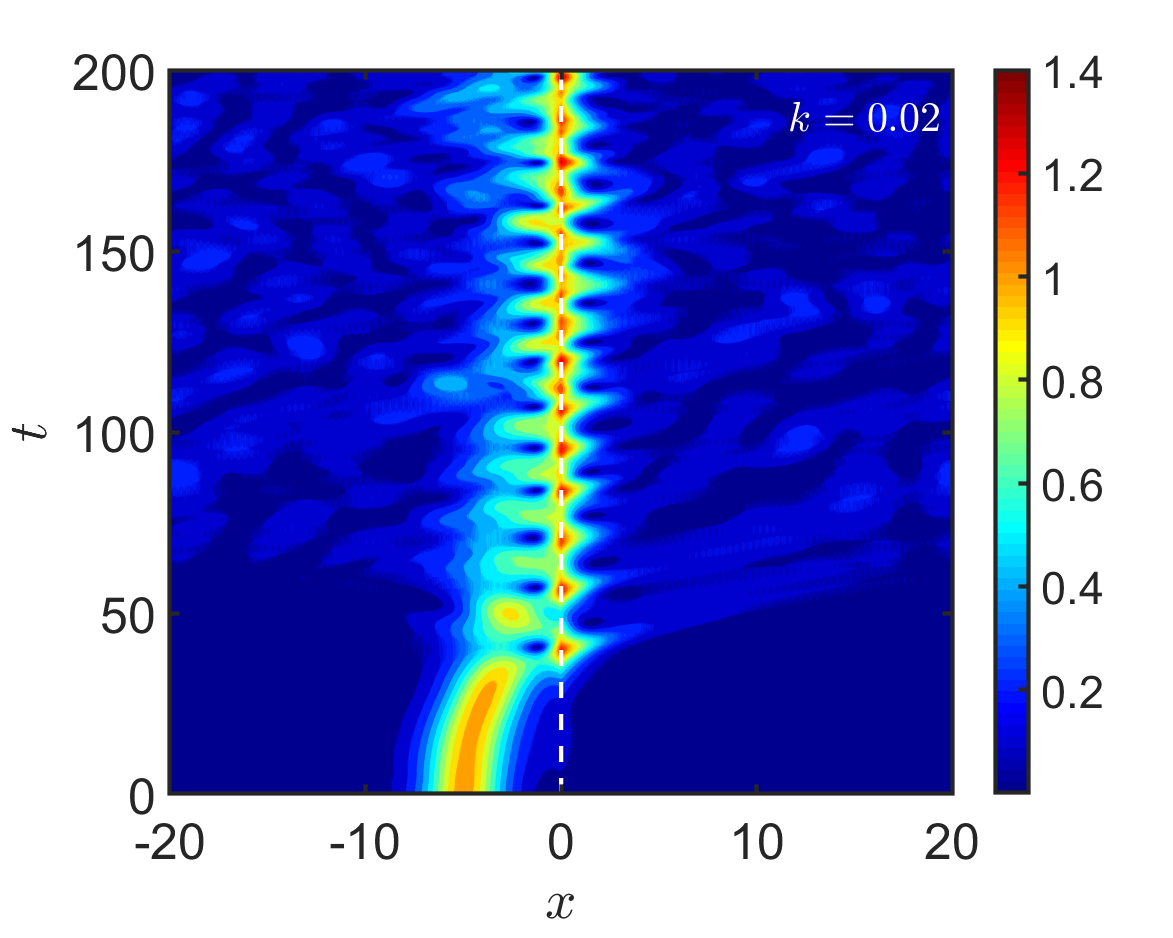}
\includegraphics[height=5cm]{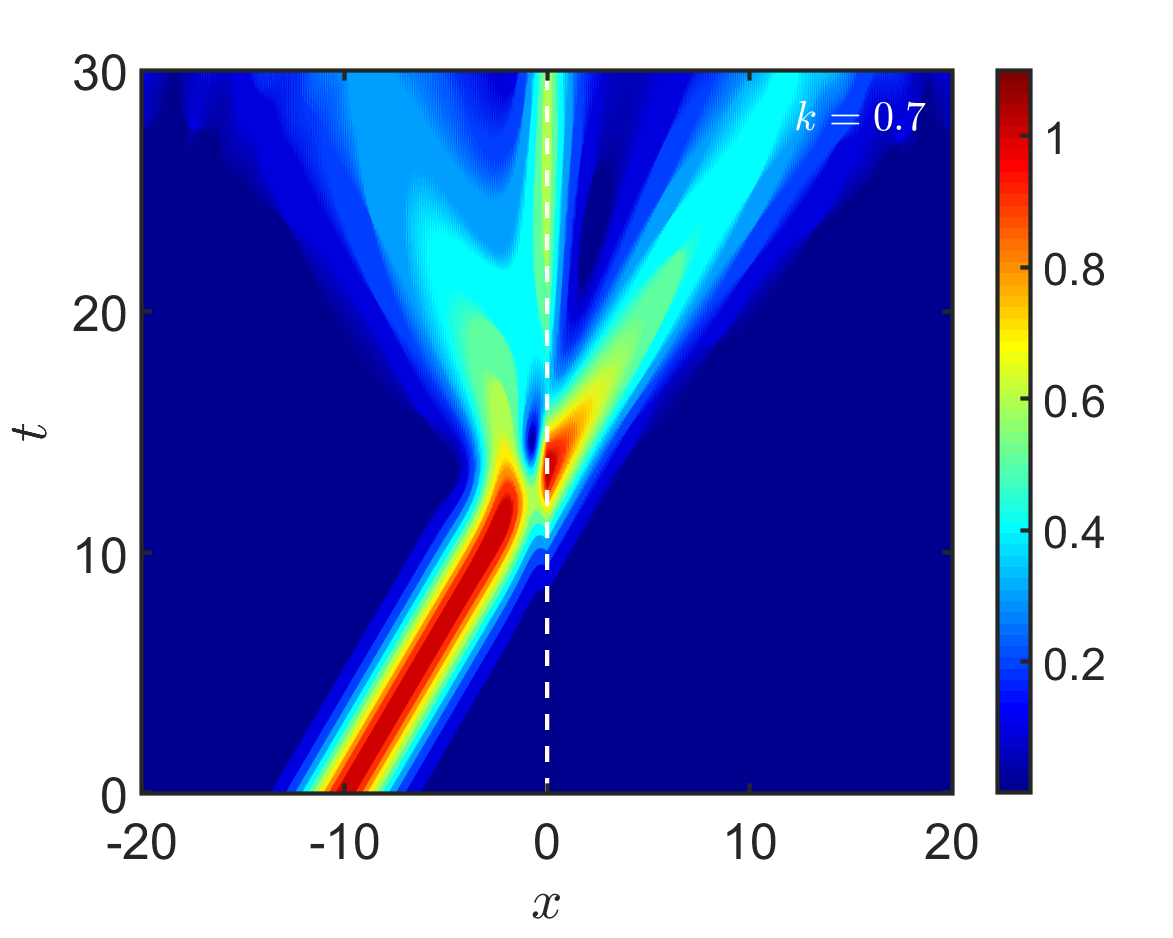}
\caption{Contour plots showing the evolution of the modulus $|u(x,t)|$ of 
a nonlocal bright soliton, initially placed at $x_0=-5$, with $k=0.02$ (left 
panel), or placed at $x_0=-10$, with $k=0.7$ (right panel); other parameter 
values are $\epsilon = 0.5$ and $d=1$, corresponding to the crossover regime, 
$\epsilon d \sim 1$. Dashed (white) line depicts the location of the point defect. 
In the left panel, the slow soliton is trapped at the location of 
the defect, along with the emission of weak radiation. 
In the right panel, the fast soliton splits into three parts, a reflected, a 
transmitted and a trapped one.}
\label{figls5}
\end{figure}

In the same regime, it is also interesting to investigate numerically 
the scattering of a moving soliton from the defect. In particular, first 
we will consider a case where the moving 
soliton features a momentum (and energy) less than that of the trapping strength,  
i.e., with $k<\epsilon$. In the left panel of Fig.~\ref{figls5}, shown is the 
outcome of such a soliton-defect interaction, for $k=0.02$ and $\epsilon=0.5$. 
It is observed that the considered slow soliton is captured by the delta potential 
due to a nonlinear resonance with the defect mode; in fact, the process involves 
small-amplitude damped oscillations of the soliton around the defect, accompanied 
by emission of radiation. On the 
other hand, for the same trap strength, $\epsilon=0.5$, if the soliton is faster, 
e.g., with $k=0.7$, then the incoming soliton splits into three parts, a reflected, 
a transmitted, and a trapped one, as shown in the right panel of Fig.~\ref{figls5}.
We have checked (results not shown here) that, similarly to the case of the local NLS 
\cite{jeremy}, depending on the value of $k$, the scenarios evolving partial 
reflection and trapping, as well as partial transmission and trapping, are possible 
too. 

\subsection{Nonlinear regime, $\epsilon d \ll 1$} 

\begin{figure}[tbp]
\centering
\includegraphics[height=5cm]{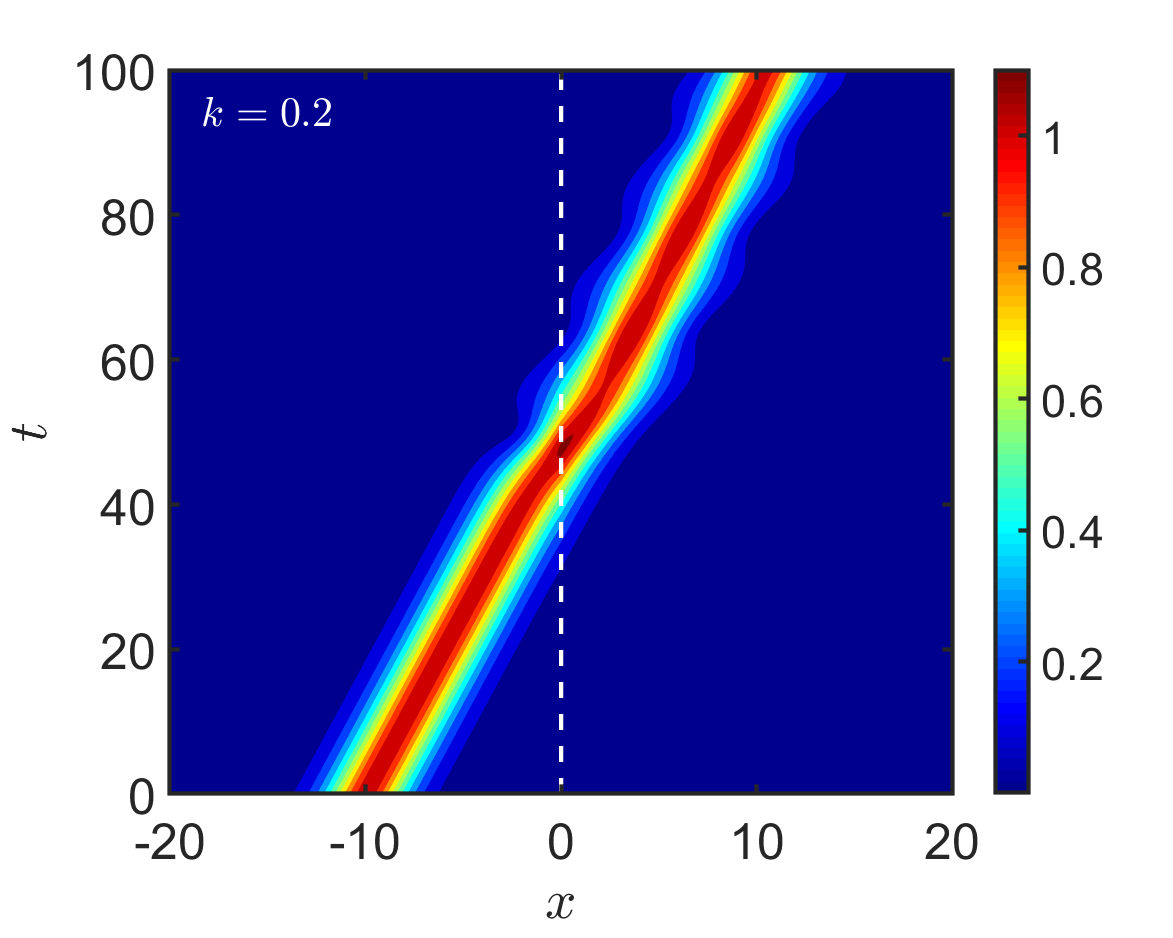}
\includegraphics[height=5cm]{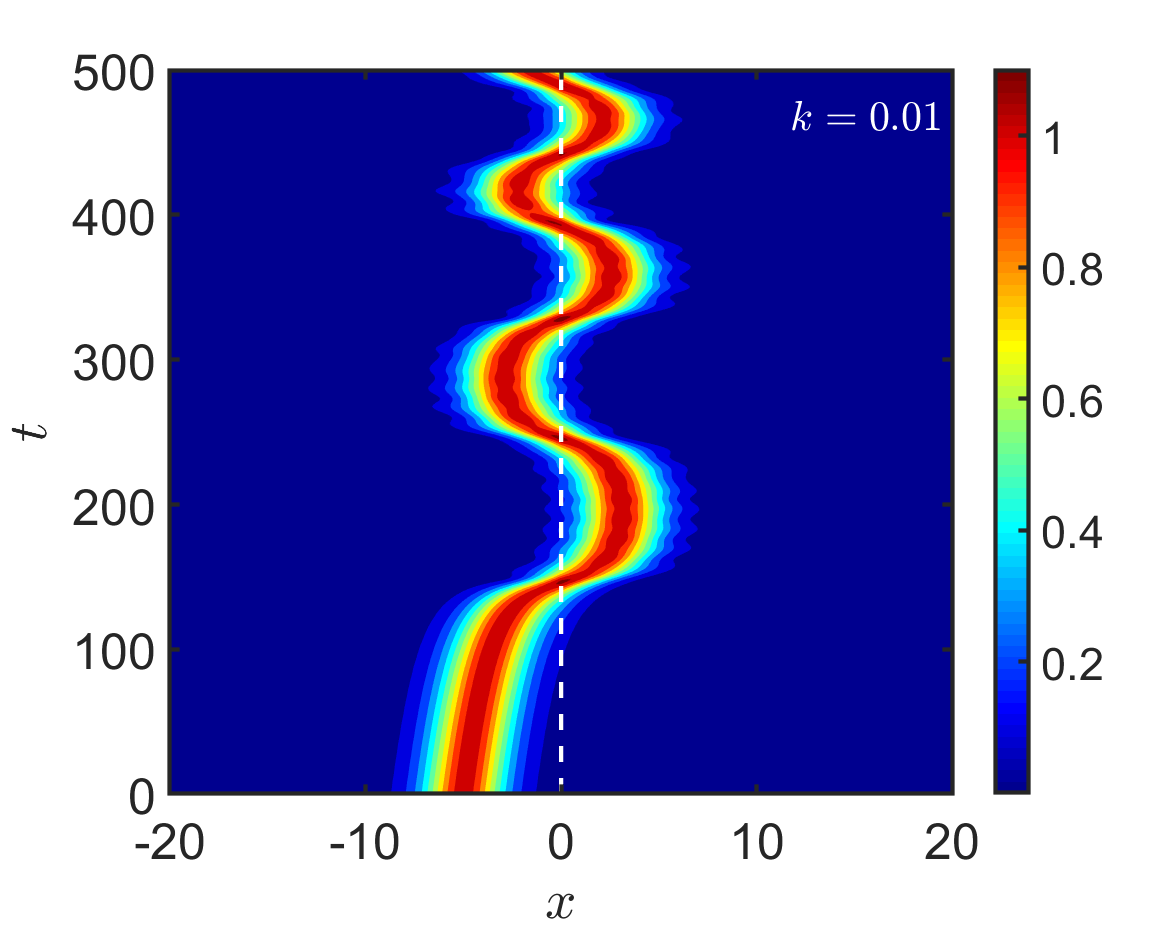}
\caption{Contour plots showing the evolution of the modulus $|u(x,t)|$ of 
a nonlocal bright soliton, with $x_0=-10$ and $k=0.2$ (left  
panel), or with $x_0=-5$ and $k=0.01$ (right panel); other parameter values 
are $\epsilon = 0.05$ and $d=1$, corresponding to the nonlinear regime, 
$\epsilon d \ll 1$. Dashed (white) line depicts the location of the point defect. 
In the left panel, the fast soliton is totally transmitted through  
the defect, while in the right panel, the slow soliton is trapped from the defect 
and performs damped oscillations around it.}
\label{fignl}
\end{figure} 

Next, we consider the nonlinear regime, where the nonlinearity 
dominates the effect of the impurity, namely $\epsilon d \ll 1$. Essentially, 
in this case, the fully nonlocal system of Eqs.~(\ref{1})-(\ref{2}), can be 
approximated by a local NLS equation incorporating the  
defect potential, or a fully nonlocal NLS in the free space.

As in the previously studied regimes, we will investigate the collision of 
a moving soliton [of the form of Eq.~(\ref{4b})] with the defect, for different 
values of wavenumber $k$. Since the nonlinear term dominates in this case, 
one should expect to observe dynamics similar to that found in the context 
of the local NLS \cite{cao,roy,jeremy}. We find that this is indeed the case, 
and there exist only two possible outcomes of the soliton-defect collision, 
namely the scenarios corresponding to $k>\epsilon$ (fast solitons) and 
$k<\epsilon$ (slow solitons). 

Relevant results are shown in Fig.~\ref{fignl} (for $d=1$ and $\epsilon=0.05$), 
where the left (right) panel depicts the fast (slow) soliton-defect interaction. 
In the left panel, the soliton is initially located at $x_0=-10$, moves towards 
the defect with $k=0.2$, and is totally transmitted through; it is observed that,  
in this case, the soliton is only slightly deformed after its transmission 
through the defect. On the other hand, in the right panel, the slow soliton with  
$k=0.01$, starts from $x_0=-5$, approaches the defect, and gets trapped, performing 
afterwards an oscillatory motion around the defect. Here, oppositely to the case 
shown in the left panel of Fig.~\ref{figls5}, the oscillations are of moderate 
amplitude (of the order of the initial soliton-defect distance). Nevertheless, 
here too, at sufficiently large times, the soliton will eventually be completely 
captured by the defect trap.     

\section{Conclusions}

Concluding, we have studied the dynamics of nonlocal bright solitons in external 
potentials. For a specific form of the nonlocal response function ---relevant to the 
evolution of optical beams in thermal media, plasmas and nematic liquid crystals---  
we have investigated two different cases: (i) external potentials that vary slowly 
over the soliton scale, and (ii) external potentials in the form of point defects. 

Starting from case (i), we have exploited the existence of an exact analytical soliton 
solution of the model, and developed an analytical approach relying on the analysis 
of the hydrodynamic form of the nonlocal NLS, and a methodology resembling the 
adiabatic perturbation theory for solitons. This way, we derived analytical nonlocal 
soliton solutions in the presence of the potential. These soliton solutions   
are exact for linear (gravitational-type) potentials, or approximate for smooth 
potentials of general form. It was shown that the center $X_0$ of the nonlocal bright 
solitons behaves like a particle, obeying the Newtonian equation of motion 
$\ddot{X_0} = -dV/dX_0$. Direct numerical simulations were found to be in very good 
agreement with the analytical predictions. In particular, in the case of the linear 
potential the soliton evolves without any deformation, following a parabolic 
trajectory, while in the case of a parabolic trapping potential, the soliton 
performs harmonic oscillations, of frequency equal to the trap frequency $\Omega$. 
In the case where the competition between the relevant spatial scales (trap 
and the soliton's widths) becomes stronger, we observed that the 
soliton oscillates featuring a distortion in its shape, but still performs 
oscillations with a frequency equal to $\Omega$. 

In case (ii), we identified three different asymptotic regimes, depending on the 
relative strength of the defect and the nonlocal nonlinearity. In the case where 
the defect dominates nonlinearity, it was found that the soliton features wave 
dynamics, being reflected by or transmitted through the defect, closely following 
the linear Schr{\"o}dinger picture. In the case where the defect and the nonlinearity 
are of equal importance, we found that the problem admits a defect mode --- the nonlinear analogue of the bound state of the linear Schr{\"o}dinger equation. 
The latter was determined analytically, as a special solution of the hydrodynamic 
equations pertaining to the nonlocal NLS model, and its existence was 
validated numerically. In the simulations we observed that the soliton-defect 
collision may lead to either complete trapping of the soliton by the defect or partial 
reflection and transmission, again accompanied by partial trapping of the soliton. On 
the other hand, in the case where the nonlinearity dominates the effect of the 
impurity, we found a behavior similar to that occurring in the local NLS equation: 
fast solitons are transmitted through the defect, while slow solitons are captured in 
the vicinity of the defect, and perform long-lived oscillations around it.        

Our analysis, which employs a combination of analytical, asymptotic, and numerical 
approaches, comes to complement previous work on the local and nonlocal NLS equations. 
We have introduced an analytical methodology, relying on the analysis of the 
hydrodynamic form of the nonlocal NLS, which may be used in a variety of relevant 
problems. This methodology allowed us to shed light on the nonlocality and the 
external potential scale competition and its concomitant role on the soliton dynamics. 
Furthermore, it should be pointed out that although the presented results are 
qualitatively similar to earlier ones corresponding to specific physical settings, 
they pertain to a broad class of physically relevant nonlocal systems, 
as mentioned above. Thus, our work offers a rather universal perspective on the 
particle-wave duality of nonlocal solitons.

It would be interesting to investigate also the dynamics of nonlocal dark solitons in 
external potentials; this is particularly challenging, as exact dark soliton solutions 
for the nonlocal NLS model are not available. Furthermore, it would be relevant 
to generalize our considerations to other localized solitary wave structures 
(such as vortices) occurring in higher-dimensional settings.

\bibliographystyle{elsarticle-num}

\begin{thebibliography}{99}

\bibitem{bec} F. Dalfovo, S. Giorgini, L. P. Pitaevskii, and S. Stringari, 
Rev. Mod. Phys. {\bf 71}, 463 (1999).

\bibitem{th1} D. Suter and T. Blasberg, Phys. Rev. A {\bf 48}, 4583 (1993).

\bibitem{th2} C. Rotschild, O. Cohen, O. Manela, M. Segev, and T. Carmon, 
Phys. Rev. Lett. {\bf 95}, 213904 (2005).

\bibitem{pl1} A. G. Litvak, JETP Lett. {\bf 4}, 230 (1966).

\bibitem{pl2} A. I. Yakimenko, Y. A. Zaliznyak, and Y. S. Kivshar, 
Phys. Rev. E {\bf 71}, 065603(R) (2005).

\bibitem{lq1} M. Peccianti and G. Assanto, Phys. Rep. {\bf 516}, 147 (2012).

\bibitem{lq2} G. Assanto, {\it Nematicons: Spatial Optical Solitons in 
Nematic Liquid Crystals} (New York, Wiley, 2012).

\bibitem{dbec} T. Lahaye, C. Menotti, L. Santos, M. Lewenstein, and T. Pfau, 
Rep. Prog. Phys. {\bf 72}, 126401 (2009).

\bibitem{kivsharagr} Y. S. Kivshar and G. P. Agrawal,
{\it Optical solitons: from fibers to photonic crystals}
(Academic Press, San Diego, 2003).

%%%%%%

\bibitem{landau} L. D. Landau and E. M. Lifschitz,
{\it Quantum Mechanics, Non-relativistic theory} (Pergamon Press, Oxford, 1965); 
%\bibitem{saku} 
J. J. Sakurai, {\it Modern Quantum Mechanics} (Addison-Wesley, 1994).

%%%%%%

\bibitem{kos} A. M. Kosevich, Physica D {\bf 41}, 253 (1981).

\bibitem{moura} M. A. de Moura, 
%Nonlinear Schrodinger solitons in the presence of an external potential
J. Phys. A: Math. Gen. {\bf 27}, 7157 (1994). %-7164

\bibitem{hulet} U. Al Khawaja, H. T. C. Stoof, R. G. Hulet, K. E. Strecker, 
and G. B. Partridge, Phys. Rev. Lett. {\bf 89}, 200404 (2002).

\bibitem{hern} C. Hernandez Tenorio, E. V. Vargas, V. N. Serkin, M. Aguero Granados, 
T. L. Belyaeva, R. P. Moreno, and L. Morales Lara, 
%Dynamics of solitons in the model of nonlinear
%Schrodinger equation with external harmonic potential, I. Bright solitons,
Quantum Electron. {\bf 35}, 778 (2005).

\bibitem{Nonlin}  R. Carretero-Gonz\'alez, D. J. Frantzeskakis, and P. G. Kevrekidis, 
Nonlinearity {\bf 21}, R139 (2008).

%%%%%%%

\bibitem{segchr} B. Alfassi, C. Rotschild, O. Manela, M. Segev, and 
D. N. Christodoulides, 
%Boundary force effects exerted on solitons in highly nonlocal nonlinear media
Opt. Lett. {\bf 32}, 154 (2007). 

%%% Also [5], [6] -- Assanto

\bibitem{natphot} G. Poy, A. J. Hess, A. J. Seracuse, 
M. Paul, S. Žumer, and I. I. Smalyukh,
%Interaction and co-assembly of optical and topological solitons. 
Nat. Phot. {\bf 16} 454 (2022). %- 461

\bibitem{assoe} M. Peccianti, A. Fratalocchi, and G. Assanto, 
%Transverse dynamics of nematicons
Opt. Express {\bf 26}, 6524 (2004). 

%%%%%

\bibitem{abdul} F. Kh. Abdullaev and V. A. Brazhnyi, J. Phys. B: At. Mol. Opt. Phys. 
{\bf 45}, 085301 (2012). 


%%%%%

\bibitem{c1} Chao-Qing Dai and Jie-Fang Zhang, Nonlinear Dyn. {\bf 100}, 1621 (2020).

\bibitem{tph} T. P. Horikis, Eur. Phys. J. Plus {\bf 135}, 562 (2020).

\bibitem{c2} Yi-Xiang Chen and Xiao Xiao, Nonlinear Dyn. {\bf 109}, 2003 (2022).

%%%%%

\bibitem{bishop} The case where the soliton width and the characteristic spatial 
scale of the potential are of the same order has also been studied for solitons 
confined in periodic potentials ---see, e.g., R. Scharf and A. R. Bishop, 
Phys. Rev. E {\bf 47}, 1375 (1993). 
 
%%%%%%

\bibitem{cao} X. D. Cao and B. A. Malomed, Phys. Lett. A {\bf 206}, 177 (1995).

\bibitem{roy} R. H. Goodman, P. J. Holmes, and M. I. Weinstein, 
Physica D {\bf 192}, 215 (2004). 

\bibitem{jeremy} J. Holmer, J. Marzuola, and M. Zworski, Commun. Math. Phys.
{\bf 274}, 187 (2007); J. Nonlinear Sci. {\bf 17}, 349 (2007).

\bibitem{assdual} C. P. Jisha, A. Alberucci, R.-K. Lee, and G. Assanto, 
%wave-particle duality
Opt. Lett. {\bf 36}, 1848 (2011).


\bibitem{natphys} M. Peccianti, A. Dyadyusha, M. Kaczmarek, and G. Assanto,
Nat. Phys. {\bf 2}, 737 (2006).

\bibitem{krolexsol} P. D. Rasmussen, O. Bang, and W. Królikowski,
%Theory of nonlocal soliton interaction in nematic liquid crystals
Phys. Rev. E {\bf 72}, 066611 (2005).


%\bibitem{zak1} V. E. Zakharov and A. B. Shabat, Sov. Phys. JETP. {\bf 34}, 62 (1972).

\bibitem{kar2} V. I. Karpman, Sov. Phys. JETP {\bf 50}, 58 (1979); 
Phys. Scr. {\bf 20}, 462 (1979).

\bibitem{kivmal} Yu. S. Kivshar and B. A. Malomed, 
Rev. Mod. Phys. {\bf 61}, 761 (1989).


%%%%%%%


\bibitem{job} W. Krolikowski, O. Bang, N. I. Nikolov, D. Neshev, J. Wyller, 
J. J. Rasmussen, and D. Edmundson, 
J. Opt. B: Quantum Semiclass. Opt. {\bf 6}, S288 (2004).

\bibitem{dum} D. Mihalache, Romanian Rep. Phys. {\bf 59}, 515 (2007).

\bibitem{bam1} B. A. Malomed, Symmetry {\bf 14}, 1565 (2022).

\bibitem{g1jpa} G. N. Koutsokostas, T. P. Horikis, P. G. Kevrekidis, and 
D. J. Frantzeskakis, J. Phys. A: Math. Theor. {\bf 54}, 085702 (2021).


%%%%%%%%%%%%%% approx soliton solutions

\bibitem{appr1} W. Krolikowski and O. Bang, Phys. Rev. E {\bf 63}, 016610 (2001).

\bibitem{appr2} N. I. Nikolov, D. Neshev, O. Bang, and W. Z. Krolikowski,
Phys. Rev. E {\bf 68}, 036614 (2003).


\end{thebibliography}

\end{document}